\documentclass[aps , pra , a4paper, twocolumn, 10pt, superscriptaddress, showpacs  ,hidelinks,showkeys, longbibliography,nofootinbib]{revtex4-2}

\usepackage[table]{xcolor}
\usepackage[hidelinks]{hyperref}
\usepackage{siunitx}
\usepackage[caption=false]{subfig}
\usepackage{amssymb, acronym, graphicx}
\usepackage{orcidlink, adjustbox}
\usepackage{lipsum, ulem}
\usepackage{graphics}
\usepackage{array, physics, float}
\usepackage{amsmath, amssymb, amsfonts, placeins}
\usepackage{color,soul}
\usepackage[capitalise]{cleveref}

\renewcommand{\vec}[1]{\mathbf{#1}}
\DeclareSIUnit{\dB}{dB}

\begin{document}

\title{Noise modelling of waveguide based squeezed light sources}
\author{E.A.T. Svanberg\orcidlink{0009-0004-4771-8295}}
\author{D. Voigt\orcidlink{0000-0001-9075-6503}}
\author{V.B. Adya\orcidlink{0000-0003-4955-6280}}
\affiliation{Department of Applied Physics, KTH Royal Institute of Technology, Roslagstullsbacken 21, Stockholm SE-106 91, Sweden}

\begin{abstract}
Squeezed states of light are used for precision metrology and quantum-enhanced measurements, with applications spanning  communication and sensing. State-of-the-art squeezed-light sources typically rely on optical cavities to achieve high, usable levels of squeezing. Recently, waveguide-based squeezed-light sources have demonstrated significant improvements in achievable squeezing, with performance currently limited by fabrication-induced losses. In this work, we present a detailed analysis of waveguide-based squeezers, examining the effects of phase noise, multiple loss mechanisms, and fundamental light leakage seeding the squeezer. We further investigate a cascaded squeezer architecture, in which a second waveguide operates as a phase-sensitive amplifier to mitigate out-coupling and detection losses. Owing to their ease of integration, robustness to high pump powers, and low intrinsic phase noise, we propose waveguide-based squeezed-light sources as a promising alternative for quantum noise reduction in future gravitational wave detectors, such as the Einstein Telescope.
\end{abstract}


\maketitle

\section{Introduction}
Squeezed states of light are engineered low-noise states with uncertainty in the phase or amplitude quadrature below the shot noise limit. They are typically generated by non-linear interactions in periodically poled (pp) crystals. Current use cases include sensitivity enhancement in high-precision metrology like gravitational wave detection\citep{aasiEnhancedSensitivityLIGO2013,abadieGravitationalWaveObservatory2011,loughFirstDemonstration62021,virgocollaborationIncreasingAstrophysicalReach2019,yuQuantumCorrelationsLight2020,tseQuantumEnhancedAdvancedLIGO2019} and bio-sensing \cite{michaelSqueezingenhancedRamanSpectroscopy2019, bowenQuantumLightMicroscopy2023, casacioQuantumenhancedNonlinearMicroscopy2021}, as well as quantum communication\cite{derkachSqueezingenhancedQuantumKey2020, usenkoContinuousvariableQuantumCommunication2025a, fletcherOverviewCVMDIQKD2025} and computing\citep{yokoyamaFullstackAnalogOptical2025, yamashimaAllopticalMeasurementdevicefreeFeedforward2025, kawasakiRealtimeObservationPicosecondtimescale2025, kashiwazakiOver8dBSqueezedLight2023}. Squeezing levels as high as \SI{15}{\decibel} have been reached in a cavity-based setup\cite{vahlbruchDetection15DB2016}. 
However, making use of these high squeezing levels is generally limited by the introduction of noise\cite{shajilal126DBSqueezed2022}, making it challenging to preserve the squeezing throughout application and detection.

State-of-the-art squeezed light sources, as used in e.g. gravitational wave detectors, use non-linear crystals, like pp-KTP\footnote[1]{potassium titanyl phosphate}, embedded into a double resonant optical cavity\cite{gaoGenerationSqueezedVacuum2024, sunSqueezingLevelStrengthened2023, zhangOptimizationSqueezingFactor2019}. 
Their disadvantage lies in its complexity and limitations in spectral squeezing performance due to the bandwidth of the cavity\cite{vahlbruchDetection15DB2016,astHighbandwidthSqueezedLight2013}. 
Especially when operated over long time periods, these squeezers show effects of gray tracking\cite{bocchini_understanding_2020, bocchiniMicroscopicOriginGray2025} and green induced infrared absorption losses\cite{shiDetectionPerfectFitting2018}, severely limiting the lifetime of the squeezer. 
Additionally, they can contribute to backscatter noise\cite{oelkerSqueezedLightAdvanced2014,chua_backscatterimpact_2014} and fluctuations of the squeezed-quadrature angle arising from alignment jitter and path-length variations in the optical cavities\cite{gaoGenerationSqueezedVacuum2024,oelkerUltralowPhaseNoise2016}. 

An alternative to the cavity-based approach is using pp-LiNb$\mathrm{O}_{3}$\footnote[2]{lithium niobate} (ppLN) waveguides as optical parametric amplifiers (OPA). Such single pass OPAs offer competitive squeezing levels owing to higher non-linearities, which result from higher confinement along longer, diffraction free, interaction lengths\cite{ kashiwazakiOver8dBSqueezedLight2023}. 
Their squeezing bandwidth is only limited by the phase matching bandwidth of the non-linear medium, which for ppLN is on the order of THz\cite{kashiwazaki_continuous-wave_2020}. 
Waveguides further offer the advantage of spatially single-mode structures and are more durable against high pump power, thus reducing gain induced diffraction effects\cite{kashiwazakiFabricationLowlossQuasisinglemode2021}. Their single-pass configuration that omits highly reflective surfaces also makes them inherently less prone to backscatter. This can ease the significant effort for optically and mechanically isolating\cite{chua_backscatterimpact_2014} squeezers in e.g. gravitational wave detectors. 
To date, squeezing levels of \SI{10}{\decibel} have been measured using ppLN waveguide sources around \SI{1545}{\nano\meter}\cite{hirotaGeneration10dBSqueezed2025}.

Currently, the out-coupling losses of the waveguide are the primary limiting factor for achieving higher squeezing levels. However, there are other factors to consider when designing a waveguide-based squeezer.
In this work we perform the first comprehensive analysis of these noise sources. This analysis can also be applied to the emerging ppLN-on-insulator (ppLNOI) platform. Its potential for miniaturisation and integration enables the effective design of fully integrated squeezed-light sources\cite{stokowskiIntegratedQuantumOptical2023,wangUltrahighefficiencyWavelengthConversion2018}. The experimental exploration\cite{nehraFewcycleVacuumSqueezing2022,mondainChipbasedSqueezingTelecom2019, lenziniIntegratedPhotonicPlatform2018,chenUltrabroadbandQuadratureSqueezing2022} of this platform has just started, further motivating our systematic in-depth analysis of expected noise sources and limitations in waveguide based squeezers.

After examining phase noise and different loss contributions, we incorporate the impact of fundamental light leakage, which is particularly relevant for on-chip systems\cite{stokowskiIntegratedQuantumOptical2023}.
We then transfer the approach of cascading squeezers for limiting noise coupling in the detection chain\cite{cavesQuantummechanicalNoiseInterferometer1981, cavesQuantumLimitsNoise1982,manceauDetectionLossTolerant2017,kwanAmplifiedSqueezedStates2024} to waveguide based systems. Our analysis presents waveguide based squeezers as an alternative to current implementations, combining comparable performance with simplified integration and improved resilience to phase noise and gain induced diffraction. This positions them as candidates for future gravitational wave detectors, including the Einstein Telescope (ET).

\section{Noise budget of a waveguide squeezed light source}
We solve the coupled mode equations for the light field quadratures inside the waveguide\cite{stokowskiIntegratedQuantumOptical2023, ouPropagationQuantumFluctuations1994}. Using the quantum field fluctuations $\delta \hat{a}$, we define the squeezed quadratures as:
 \begin{align}
     & \delta\hat{X}_+ = \delta \hat{a} + \delta\hat{a}^\dagger \label{eq:AmpQuadrature}\\
     & \delta\hat{X}_- = i(\delta\hat{a}^\dagger  - \delta\hat{a}\label{eq:PhaseQuadrature})
 \end{align}
Here the $-$ and $+$ are used to indicate squeezing in the phase quadrature and anti-squeezing in amplitude quadrature.
This leads to the variances of the respective quadratures being:
\begin{align} \label{eq:SimpleSqueeze}
    &V_- = \exp({-2 \sqrt{\alpha_{\text{OPA}} P_{\text{SHG}}}})  = \exp(-2R)\\ 
    &V_+ = \exp({2 \sqrt{\alpha_{\text{OPA}} P_{\text{SHG}}}}) = \exp(2R)
\end{align}
in which we define the squeezing parameter as $R=\sqrt{\alpha_{\text{OPA}} P_{\text{SHG}}}$.  $P_{\text{SHG}}$ is the power of the second harmonic generation (SHG) light used to pump the OPA waveguide and $\alpha_{\text{OPA}}$ is the conversion efficiency in units of $W^{-1}$. 

Squeezed light is usually detected using balanced homodyne detection where a local oscillator (LO) is used to optically amplify the weak signal whilst its own noise is removed by common mode rejection. Tuning the relative phase $\theta$ of the LO allows to determine which quadrature is measured. The inability to precisely match the LO and squeezed quadrature introduces phase noise. This is one of the two primary factors limiting squeezing levels, with the other one being losses.
Losses can be understood as adding vacuum fluctuations into the (anti-)squeezed quadratures and is modelled as a partially transmissive mirror. 
If we define the efficiency or transmission of the mirror as $\eta$ and consider the LO phase $\hat{\theta} =  \theta + \delta\hat{\theta}$ with small fluctuations $\langle sin^2(\delta\hat{\theta})\rangle =\theta_{\text{rms}}^2 \ll 1$, the measured variance becomes\cite{mccullerLIGOsQuantumResponse2021}:
\begin{align}
    &V_{\text{meas}} = \eta [1-\theta_{\text{rms}}^2](\cos^2(\theta) V_+ + \sin^2(\theta) V_-) \\ & + \eta \theta_{\text{rms}}^2(\sin^2(\theta) V_+ + \cos^2(\theta) V_-) + (1-\eta)\nonumber
\end{align}
Since $V_-\!\leq\!1\!\leq\!V_+$, meaning the squeezed noise is reduced below the vacuum fluctuations, both loss and phase noise have a greater effect on squeezing than on anti-squeezing. Loss decreases the achievable squeezing, keeping the squeezing level above the $1\!-\!\eta$ threshold. Phase noise, $\theta_\text{rms}$, couples anti-squeezing into the measured squeezing quadrature and vice versa. 

\begin{figure}[htb!]
  \centering
  \includegraphics[width=1\columnwidth]{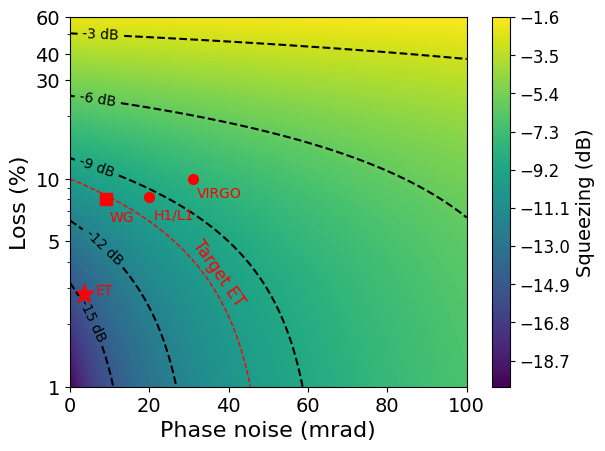}
  \caption{A contour plot of the maximum squeezing as a function of loss and phase noise. The red dots show current squeezers used at the LIGO and VIRGO detectors\cite{LIGOSqueezers, Virgosqueezer}, accounting only for injection losses. The red dashed line gives current Einstein Telescope (ET) requirements of effective squeezing\cite{ETDocs}, whilst the star is a suggested squeezer for ET\cite{ETDocs, ETSqueezer}. The red square is the current best waveguide squeezer\cite{hirotaGeneration10dBSqueezed2025}. At lower squeezing levels, losses are the dominant limiting factor, while at higher squeezing levels, minimizing phase noise becomes equally critical.}
  \label{Fig:Contour}
\end{figure}

Due to this coupling, increasing power also amplifies the coupled anti-squeezing, thereby reducing the measurable squeezing levels. Thus, for each level of phase noise, there exists an optimal squeezing parameter, $R_{\text{opt}}$, for maximizing the achievable squeezing. Regardless of the loss in the system, this is given by:
\begin{equation}
    R_{\text{opt}} =  -\frac{1}{4}\ln(\frac{\theta_{\text{rms}}^2}{1-\theta_{\text{rms}}^2}) \approx -\frac{1}{2}\ln(\theta_{\text{rms}}) 
\end{equation}
Hence, for a given squeezing efficiency $\alpha_{\text{OPA}}$ and phase noise $\theta_{\text{rms}}$, the optimal (SHG) power $P_{\text{SHG,opt}}$ to maximize the squeezing performance is: 
\begin{equation}
    P_{\text{SHG,opt}} = \frac{1}{4\alpha_{\text{OPA}}} \ln(\theta_{\text{rms}})^2.
\end{equation}
With this, we get a maximum squeezing level of:
\begin{equation}
    V^{\text{min}}_- = 2 \eta \theta_{\text{rms}} \sqrt{1\!-\!\theta_{\text{rms}}^2} + 1\!-\!\eta \approx 2 \eta \theta_{\text{rms}} + 1\!-\!\eta.
\end{equation}
This maximum squeezing as a function of loss and phase noise is shown in figure \ref{Fig:Contour}. The parameters used in all plots, unless stated otherwise, are chosen to reflect realistic parameters taken from \cite{kashiwazakiOver8dBSqueezedLight2023}. 

Next, we examine sources of phase noise, different types of loss, and discuss how leakage of the unconverted pump after the SHG stage limits detectable squeezing.

\label{section: noise_budget}
\subsection{Phase noise and loss considerations}
\label{phase_noise}
Phase noise, sometimes called phase jitter, is observed if quadrature fluctuations occur faster than the measurement time. From figure \ref{Fig:Contour} we see that achieving high squeezing requires not only low loss but also low phase noise. The figure includes examples of current state-of-the-art squeezed light sources. The sources for VIRGO and LIGO only include injection loss into the interferometer and are at $1064$ nm\cite{Virgosqueezer, LIGOSqueezers}. The red dashed line is the targeted 10 dB effective squeezing for the ET\cite{ETDocs} and the star is a suggested squeezed light source at $1550$ nm for ET\cite{ETSqueezer}. The currently best waveguide squeezer around $1550$ nm\cite{hirotaGeneration10dBSqueezed2025} is indicated by the red square, sitting right below the target line for ET. Although this does not include interferometer losses, the ease of integration and intrinsically low phase noise of waveguide-based squeezed light sources make them a promising alternative for ET. This statement is further substantiated by the experimental reduction of phase noise from 14 mrad to 9 mrad, simply by changing the phase control scheme\cite{hirotaGeneration10dBSqueezed2025}.

Experimentally, it is challenging to separate the effects of optical loss and inherent phase noise. To do so, high levels of pump power should be used to properly show the phase noise distinctive reduction in squeezing above the optimal power. 
 
In cavity based squeezers, phase noise contributions include: cavity length fluctuations, temperature fluctuations of the non-linear crystal, amplitude and phase noise from the pump, and sensing noise of various locking fields\cite{dwyerQuantumNoiseReduction2013}. The latter are needed to control the optical path lengths and the squeezing quadrature\cite{oelkerUltralowPhaseNoise2016, vahlbruchCoherentControlVacuum2006}. High phase noise from the pump is especially critical in this case as it leads to time dependent mixing of the quadratures inside the cavity. Outside the cavity, this effect can be mitigated by locking the detection quadrature to the squeezing. However, intra-cavity mixing cannot be mitigated by external quadrature locking. Avoiding the cavity infrastructure by using a waveguide eliminates these topology-specific phase noise contributions. Besides reducing pump phase and amplitude noise contributions, it also removes cavity length fluctuations and requires fewer locking fields, further reducing phase noise sources.

ppLN waveguides are known to suffer from photorefraction, which can distort the spectrum of SHG and lead to wavelength deviations of the squeezed vacuum. However, these effects can be counteracted by doping\cite{kong_recent_2020}, heating\cite{mondain_photorefractive_2020} and adaptive poling\cite{chen_adapted_2024,wang_impact_2025} of the waveguide. 

\subsection{Effects of losses}

\subsubsection{In-coupling and out-coupling losses}
\begin{figure}[htb!]
  \centering
  \includegraphics[width=1\columnwidth]{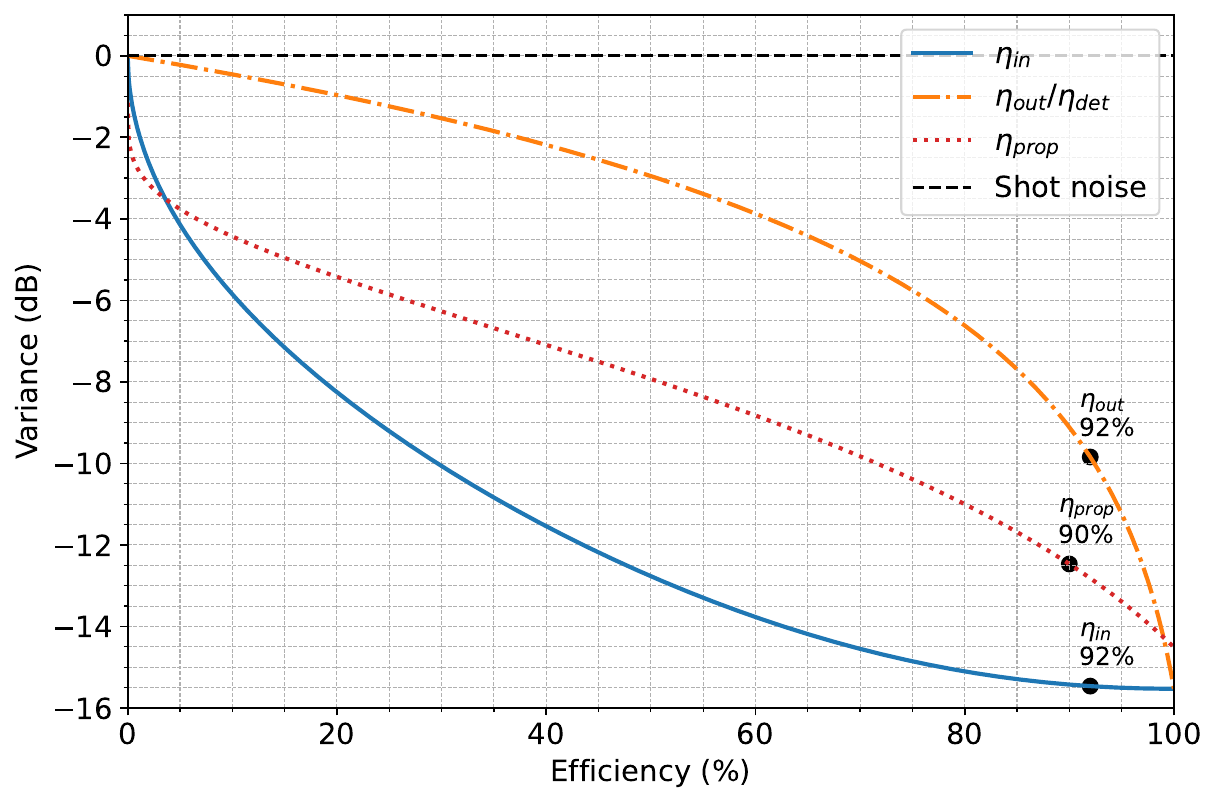}
  \caption{Plot showing the effects of in-coupling, out-coupling, detection, and propagation efficiencies on squeezing. Out-coupling and detection losses have an identical effect. Black dots indicate typical values of out-coupling, propagation, and in-coupling losses from \cite{kashiwazakiOver8dBSqueezedLight2023}. }
  \label{Fig:4panel}
\end{figure}

The in-coupling and out-coupling loss are important parameters of the waveguide. Both affect the achievable squeezing levels, as shown in figure \ref{Fig:4panel}. In-coupling loss can easily be mitigated by increasing the pump power. Out-coupling loss, however, provides a limit to the maximum squeezing achievable and is challenging to mitigate without using advanced techniques such as the cascaded squeezing described in section~\ref{section:double_sqzer}. 

\subsubsection{Propagation losses}
An important loss related to the waveguide's manufacturing is propagation loss, caused by surface roughness, intrinsic imperfections, or contamination of the material. Its main effect is scattering into higher order modes, which reduces squeezing in the fundamental mode. We model propagation loss and the squeezing parameter as constant throughout the waveguide. Considering a continuous infinitesimal coupling of vacuum states at each point, the output will become\cite{yamashimaAllopticalMeasurementdevicefreeFeedforward2025}:
\begin{align}
    &\delta \hat{X}_+^{\text{out}} = e^{(r-\gamma)L} \delta \hat{X}_+^{\text{in}} + \sqrt{\frac{\gamma(e^{2(r-\gamma)L} -1)}{r-\gamma} } \hat{X}^{\text{vac}}\\
    &\delta \hat{X}_-^{\text{out}} = e^{-(r+\gamma)L} \delta \hat{X}_-^{\text{in}} + \sqrt{\frac{\gamma(1- e^{-2(r+\gamma)L} )}{r+\gamma} } \hat{X}^{\text{vac}}
\end{align}
where $r=R/L$ and $\gamma = \frac{\Gamma}{L}$ are the squeezing factor and loss per unit length respectively. 

This leads to a different effective total propagation efficiency for squeezing and anti-squeezing: 
\begin{align}
    &\eta_{\text{prop,asqz}}' = \frac{R(e^{2(R-\Gamma)} - 1)}{(R-\Gamma)(e^{2R}-1)} \\
    &\eta_{\text{prop,sqz}}' = \frac{R(1-e^{-2(R+\Gamma)})}{(R+\Gamma)(1-e^{-2R})}
\end{align}

In figure \ref{Fig:4panel}, we show the squeezing as a function of total propagation efficiency, $\eta_{\text{prop}} = e^{-2\Gamma}$. This is different from the effective propagation efficiency, because gradual loss is less detrimental to squeezing than the same loss clumped at the output. In the case of anti-squeezing, if we assume $R\gg\Gamma$, this leads to $\eta_{\text{prop,asqz}}' \approx \eta_{\text{prop}}$. However, the same does not hold for squeezing. 

\subsubsection{Detection losses} 
\begin{figure*}[htb!]
  \centering
  \includegraphics[width=0.8\textwidth]{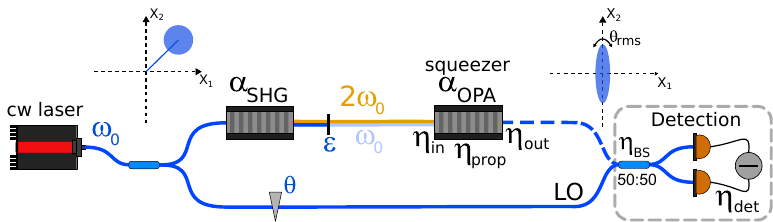}
  \caption{Sketch of the integrated squeezer architecture with the two waveguides connected by fibre to the continuous wave (CW) source and detection. Highlighted are the key analysis parameters described in the text: $\alpha_{\text{SHG}}$, $\alpha_{\text{OPA}}$, $\epsilon$, $\eta_{\text{in}}$, $\eta_{\text{prop}}$, $\eta_{\text{det}}$, $\eta_{\text{out}}$, $\eta_{\text{BS}}$, $\theta_{\text{rms}}$.}
  \label{Fig:SqueezerArchitecture}
\end{figure*}

The balanced photodetectors for detection of squeezed light use high quantum efficiency photodiodes with low noise electronic circuitries. This ensures that the overall photodetector electronic noise, dominated primarily by Johnson noise, stays well below the shot noise level. The photodetectors quantum efficiency and the mode overlap between LO and the squeezed field are the key aspects of detection loss. Any fluctuation in spatial mode overlap creates a pathway for coupling anti-squeezing into squeezing, thereby reducing the overall measured squeezing levels. To improve mode overlap, shaping the local oscillator via a secondary waveguide\cite{roman-rodriguezMultimodeSqueezedState2024} or using all-fiberised setups\cite{nakamuraLongtermStabilitySqueezed2024} is beneficial. An example of such a setup is shown in figure \ref{Fig:SqueezerArchitecture}, where mode overlap losses can be minimized by implementing the detection using single-mode fibres.

\subsection{Leakage of unconverted fundamental field}
\label{leakage}
The fundamental light used for driving the SHG process can leak into and through the OPA system. It can then act as a LO for the vacuum fluctuations in the path intended to be the LO. 
Therefore, vacuum fluctuations are measured alongside the squeezed light, degrading the quality of the observed squeezing.

The LO path has an average amplitude $\Bar{\beta}$ ($P_{LO} = |\Bar{\beta}|^2$) with fluctuations $\delta \hat{\beta}$, whilst the OPA path has fluctuations $\delta \hat{a}$ and an average amplitude $\sqrt{\epsilon} \Bar{\beta}$, where $\epsilon$ is the ratio of power in the OPA path to the LO path. For completeness we also consider the effects of an imperfect 50/50 beamsplitter (BS). The BS has a transmission $\eta_{BS}$ and reflectivity $1-\eta_{BS}$. The output fields after the BS are:
\begin{align}
    &\hat{C} = \sqrt{\eta_{BS}} \hat{A} + \sqrt{1-\eta_{BS}} \hat{\beta}e^{i\theta} \\
    &\hat{D} = \sqrt{1-\eta_{BS}} \hat{A} - \sqrt{\eta_{BS}} \hat{\beta}e^{i\theta}
\end{align}
where $\hat{A} = \sqrt{\epsilon} \Bar{\beta} + \delta \hat{a}$, $\hat{\beta} = \Bar{\beta} + \delta \hat{\beta}$, and $\theta$ is the LO phase chosen such that $\Bar{\beta}^* = \Bar{\beta}$. The generated photocurrents are proportional to $\hat{C}^\dagger \hat{C}$ and $\hat{D}^\dagger \hat{D}$, this gives the differential photocurrent signal as:
\begin{align}
    &i_- \propto \hat{C}^\dagger \hat{C}- \hat{D}^\dagger \hat{D} = (2\eta_{BS}-1)(\epsilon |\Bar{\beta}|^2 + \sqrt{\epsilon}\Bar{\beta}\delta \hat{X}_+^a)  \\ &+  \Bigl( (1-2\eta_{BS}) (|\Bar{\beta}|^2 +  \Bar{\beta}\delta \hat{X}_+^\beta)\Bigl) + \Bigl( 2\sqrt{\eta_{BS}(1-\eta_{BS})}  \nonumber\\ & \cdot \Bigr[2\sqrt{\epsilon} |\Bar{\beta}|^2 \cos(\theta) + \sqrt{\epsilon} \Bar{\beta}(\delta\hat{X}_+^\beta \cos(\theta) + \delta\hat{X}_-^\beta \sin(\theta))  \nonumber\\ 
    & + \Bar{\beta} (\delta\hat{X}_+^a \cos(\theta) + \delta\hat{X}_-^a \sin(\theta))\Bigr] \Bigl) \nonumber
\end{align}
Terms proportional to $\delta \hat{a}^2$  and $\delta \hat{\beta}^2$ have been neglected under the assumption that the LO power is much greater than the vacuum fluctuations ($\Bar{\beta} \gg \delta \hat{a}, \delta \hat{\beta}$). 

When measuring squeezing with a spectrum analyser, we examine the frequency spectrum of the RMS voltage noise, which is proportional to the Fourier transform of the photocurrent variance. Assuming only vacuum fluctuations in the LO path, the measured photocurrent variance is given by:
\begin{align}
    &\label{eq:BSImperfect} V_{meas}  \propto  \Bigl( 4 \eta_{BS}(1- \eta_{BS})[V_+ \cos^2(\theta) + V_-\sin^2(\theta)\\ & + \epsilon] \Bigl) P_{LO} 
     + \Bigl((2 \eta_{BS}-1)^2[1+\epsilon V_+]\Bigl) P_{LO}  +   \nonumber\\ & \Bigl(2 \sqrt{\eta_{BS}(1-\eta_{BS})}(2\eta_{BS}-1)\sqrt{\epsilon}  \cos(\theta)[V_+-1] \Bigl) P_{LO} \nonumber
\end{align}
By fully blocking the OPA path we can directly measure shot noise and eliminate the impact of leakage. Consequently, we normalize the measurement to shot noise by dividing by $P_{LO}$. 

\begin{figure}
    \centering
    \includegraphics[width=0.8\linewidth]{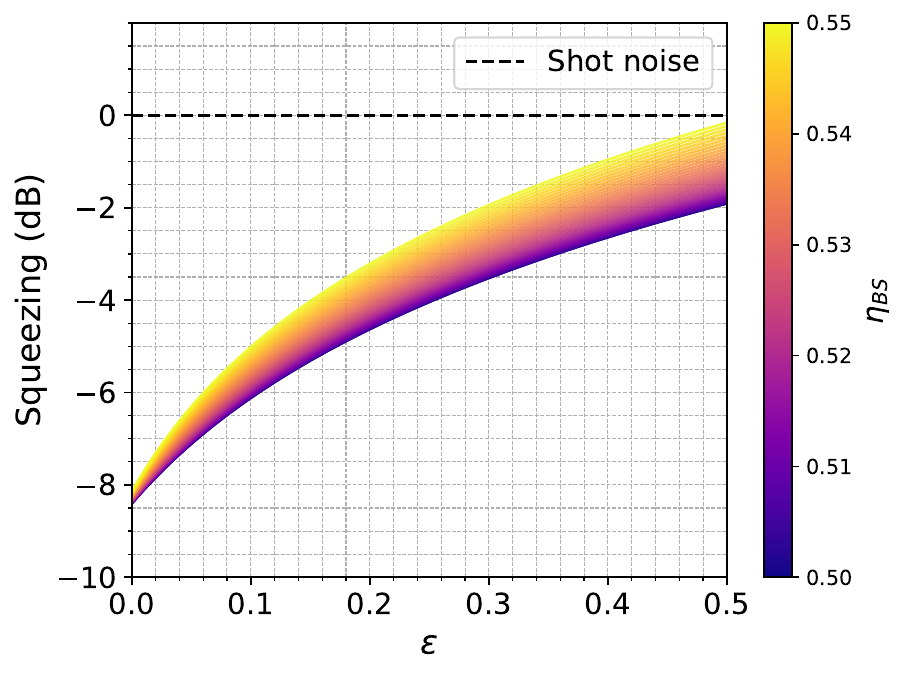}
    \caption{Squeezing as a function of leakage level, $\epsilon$, for varying levels of BS ratios. The ideal case is for a 50/50 BS and the squeezing is worsened for greater leakage levels and imperfect BS ratios.}
    \label{fig:SqzVSBS}
\end{figure}

Figure \ref{fig:SqzVSBS} illustrates the effect of increasing leakage on the measured squeezing for different BS ratios, assuming initial squeezing and anti-squeezing levels of 8.4 dB and 18 dB respectively. Leakage reduces the observed squeezing while slightly increasing the measured anti-squeezing, as it introduces vacuum fluctuations in the LO path that are not included in the shot-noise calibration. At low leakage levels, the squeezing remains relatively insensitive to small BS imperfections. However, at higher leakage and squeezing levels, the system becomes increasingly sensitive to BS imperfections, which introduce additional anti-squeezing proportional to the leakage level, as described by Eq.~\ref{eq:BSImperfect}.

In this analysis we have assumed that the common mode rejection is left unchanged with varying BS ratios. A non-50/50 BS can still be used to get high common mode rejection ratio, however the photodiode output signals will have to be matched electronically.

Notably, standard single-mode fibres at the SHG wavelength can be used to filter residual fundamental leakage, enabling compact integration without additional optical elements. In free-space implementations, care must be taken to prevent seeding. The same applies to integrated squeezers, where appropriate on-chip filtering is required. For leakage into a gravitational wave detector, the effect is even larger and must be eliminated\cite{dwyerQuantumNoiseReduction2013}. 

\section{Mitigating out-coupling loss and improving signal-to-noise ratio}
\label{section:double_sqzer}

Phase-sensitive amplification of squeezed states has been proposed theoretically\cite{cavesQuantummechanicalNoiseInterferometer1981, cavesQuantumLimitsNoise1982} and demonstrated experimentally\cite{manceauDetectionLossTolerant2017, nehraFewcycleVacuumSqueezing2022, kwanLosstolerantDetectionSqueezed2025} as an effective solution for mitigating the impact of losses in the detection chain. In a recent work, the cascading of two cavity-based OPAs was investigated to mitigate photodiode losses, with a focus on gravitational wave detectors operating at 2 $\mu m$ wavelengths\cite{kwanAmplifiedSqueezedStates2024}. This architecture can also be used to improve signal-to-noise ratio in biosensing, such as in squeezing enhanced Raman spectroscopy\cite{michaelSqueezingenhancedRamanSpectroscopy2019}. In general, this setup is known as an SU(1,1) interferometer\cite{ouQuantumSU11Interferometers2020}. Practical implementation for cascaded phase-sensitive amplifiers for gravitational wave detectors has been discussed in \cite{kwanLosstolerantDetectionSqueezed2025} and an identical scheme using waveguides is presented here. 

By applying this technique, all types of losses occurring after the amplification step can be  mitigated. Due to the high out-coupling losses typically associated with waveguides, particularly in on-chip systems, this method has been used to extend detection bandwidth and reduce the effects of waveguide out-coupling losses\cite{nehraFewcycleVacuumSqueezing2022}. This approach has previously enabled the measurement of squeezing at 1.7 dB across a 55 THz bandwidth, despite more than 50 \% losses\cite{shakedLiftingBandwidthLimit2018}. 

Although the technique has been widely used, a comprehensive model for waveguides has not yet been shown. Here we follow the working of Kwan et.al.\cite{kwanAmplifiedSqueezedStates2024} to model the phase sensitive amplification of squeezed states. We incorporate conventional loss terms and also analyse phase noise contributions from both waveguide OPAs.

\begin{figure}[htb!]
  \centering
  \includegraphics[width=1\columnwidth]{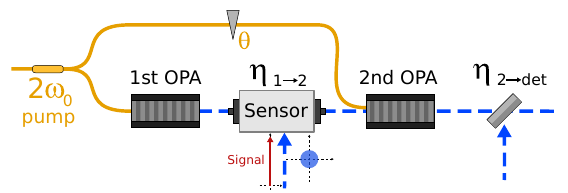}
  \caption{A schematic of the cascaded squeezing architecture, which can be used to mitigate losses after the sensor using a second OPA to amplify the signal and vacuum state.}
  \label{Fig:CascadedSqueezingFigure}
\end{figure}

A schematic of a cascaded squeezing setup is shown in figure \ref{Fig:CascadedSqueezingFigure}. Both waveguide OPAs are pumped with the same frequency pump. The initial squeezed state after the first OPA becomes anti-squeezed, along with the vacuum fluctuations, by the second OPA. The squeezing is then measured with respect to these vacuum fluctuations, allowing losses after the second OPA to become suppressed. The efficiency 
\begin{equation}
    \eta_{2\rightarrow\text{det}} = \eta_{2,\text{prop}}\eta_{2,\text{out}} \eta_{\text{det}}
\end{equation} 
accounts for the effect of all losses after the second OPA, including out-coupling and propagation within the second OPA, as well as detection loss. Likewise
\begin{equation}
    \eta_{1\rightarrow 2} = \eta_{1,\text{prop}} \eta_{1,\text{out}} \eta_{\text{sensor}}\eta_{2,\text{in}}
\end{equation} 
represents losses between the first and second OPA. This includes out-coupling and propagation loss within  the first OPA, the loss of the sensor (such as injection to a gravitational wave detector), and in-coupling to the second OPA. 

\subsection{Effective efficiency}
\begin{figure*}[htb!]
  \includegraphics[width=0.45\textwidth]{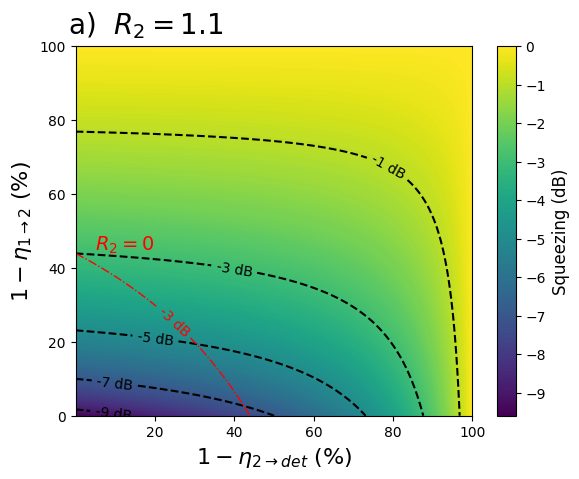}
  \includegraphics[width=0.45\textwidth]{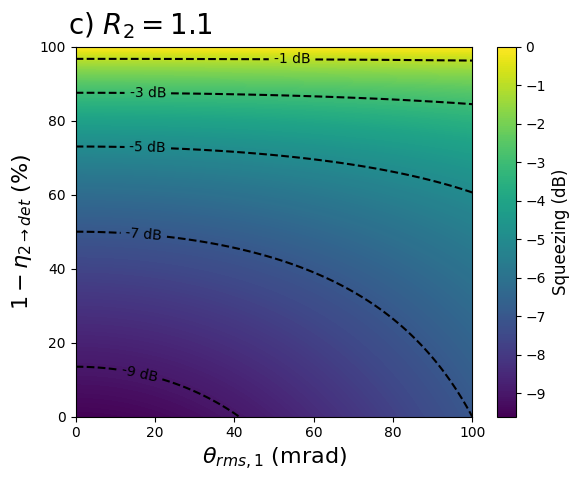}
  \hfill
  \includegraphics[width=0.45\textwidth]{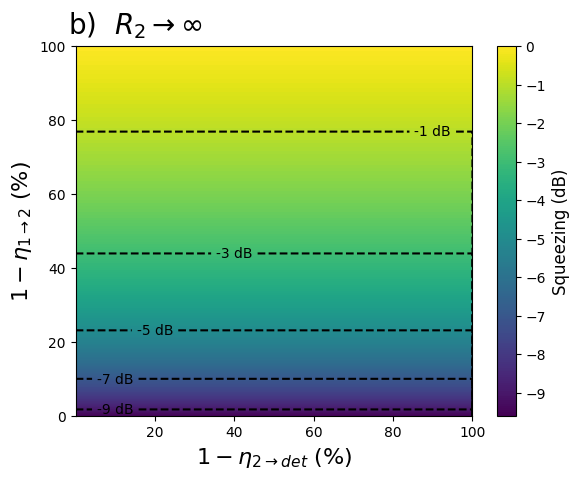}
  \includegraphics[width=0.45\textwidth]{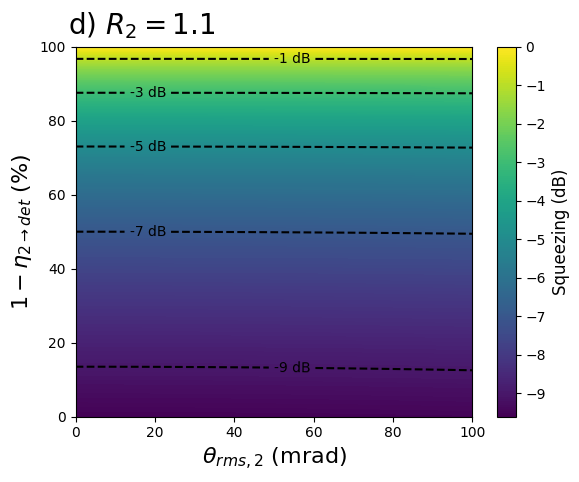}
\caption{Contour plots of  Effective squeezing for different values of $\eta_{1\rightarrow 2}$, $\eta_{2\rightarrow \text{det}}$, $\theta_\text{rms,1}$, $\theta_\text{rms,2}$, and $R_2$. $R_1 = 1.1$ in all scenarios. a)  Effective squeezing levels versus  $\eta_{1\rightarrow 2}$ and $\eta_{2\rightarrow \text{det}}$ for $R_2=1.1$. The red line is the 3 dB line for $R_2 = 0$, meaning without the effect of the second OPA. b)  Effective squeezing levels versus  $\eta_{1\rightarrow 2}$ and $\eta_{2\rightarrow \text{det}}$ for $R_2\rightarrow \infty$.  c)    Effective squeezing levels versus  $\eta_{2\rightarrow \text{det}}$ and $\theta_\text{rms,1}$ for $R_2=1.1$. d)  Effective squeezing levels versus  $\eta_{2\rightarrow \text{det}}$ and $\theta_\text{rms,2}$ for $R_2=1.1$. }
\label{fig:ContourCascade}
\end{figure*}

We use a matrix formalism to calculate the output variance from the cascaded squeezers. The details are found in appendix \ref{app:Matrix}. From this, the amplified phase and amplitude variance without phase noise becomes:
\begin{align}\label{eq:VMinusAmp}
    & V^{\text{amp}}_- =  \eta_{1\rightarrow 2}\eta_{2\rightarrow \text{det}} e^{-2R_1 + 2R_2} +
    \\&  \eta_{2\rightarrow \text{det}} (1-\eta_{1\rightarrow 2}) e^{2R_2} + 1-\eta_{2\rightarrow \text{det}}  \nonumber \\
    & V^{\text{amp}}_+ =  \eta_{1\rightarrow 2}\eta_{2\rightarrow \text{det}} e^{2R_1 - 2R_2}  + \label{eq:VPlusAmp} 
    \\&  \eta_{2\rightarrow \text{det}} (1-\eta_{1\rightarrow 2}) e^{-2R_2} + 1-\eta_{2\rightarrow \text{det}}  \nonumber
\end{align}
where $R_1$ and $R_2$ are the squeezing parameter of the first and second squeezer respectively. 

The effective measurable squeezing is calculated by dividing the amplified variance with the amplified vacuum:
\begin{equation}
    V^{\text{eff}}_- = \frac{V^{\text{amp}}_-}{V^{\text{amp}}_-|_{R_1 = 0}} = \eta_{1\rightarrow 2} \eta_{\text{eff}} e^{-2R_1} + 1- \eta_{1\rightarrow 2} \eta_{\text{eff}}
\end{equation}
where we have defined an effective efficiency after the second OPA:
\begin{equation} \label{eq:EffectiveEfficiency}
    \eta_{\text{eff}} = \frac{\eta_{2\rightarrow \text{det}}}{\eta_{2\rightarrow \text{det}} + (1-\eta_{2\rightarrow \text{det}})e^{-2R_2}}
\end{equation}

Panel a and b of figure \ref{fig:ContourCascade} show the effective squeezing as a function of $\eta_{1\rightarrow 2}$ and $\eta_{2 \rightarrow \text{det}}$ for different $R_2$. To generate roughly 10 dB of initial squeezing, $R_1=1.1$ was used. While losses after the second OPA can be mitigated, losses occurring between the two OPAs establish a fundamental limit on the achievable squeezing. In cases of substantial out-coupling or detection losses, anti-squeezing from the second OPA can significantly enhance the measurable squeezing; however, it cannot increase the squeezing produced by the first OPA.

\subsection{The effect of phase noise}
We now consider the effects of phase noise with $\theta_\text{rms,1}$ and $\theta_\text{rms,2}$ being the phase noise of the first and second squeezer, respectively. Setting $\theta_\text{rms,2} = 0$ and only considering the effects of $\theta_\text{rms,1}$, we can write the new effective squeezing as: 
\begin{align}
   & V^{\text{eff}}_-(\theta_\text{rms,1}) = \eta_{1\rightarrow 2} \eta_{\text{eff}} e^{-2R_1} \cos^2(\theta_\text{rms,1}) + \notag \\
   &\eta_{1\rightarrow 2} \eta_{\text{eff}} e^{2R_1} \sin^2(\theta_\text{rms,1}) + 1- \eta_{1\rightarrow 2} \eta_{\text{eff}} \label{eq:VeffTheta1}
\end{align}
The effect of phase noise in the first OPA couples part of the anti-squeezing to the squeezing in an identical way as for a single OPA.

The situation when considering $\theta_\text{rms,2}$ becomes more complicated. Using equations \ref{eq:VMinusAmp} and \ref{eq:VPlusAmp} and setting $\theta_\text{rms,1}=0$ the effective squeezing becomes: 
\begin{align} \label{eq:V_minusEff}
   & V^{\text{eff}}_-(\theta_\text{rms,2}) = \\ & \nonumber \frac{V^{\text{amp}}_- \cos^2(\theta_\text{rms,2}) + V^{\text{amp}}_+ \sin^2(\theta_\text{rms,2})}{[V^{\text{amp}}_- \cos^2(\theta_\text{rms,2}) + V^{\text{amp}}_+ \sin^2(\theta_\text{rms,2})]|_{R_1=0}}
\end{align}
By looking at the form of equation \ref{eq:VMinusAmp} and \ref{eq:VPlusAmp} we see that if $R_2\geq R_1$,  then $V^{\text{amp}}_- \geq V^{\text{amp}}_+$. Since $\theta_\text{rms}$ is small, this implies that $V^{\text{eff}}_-(\theta_\text{rms,2}) \approx V^{\text{amp}}_-/(V^{\text{amp}}_-|)_{R_1=0}$. Thus the effective squeezing is largely unaffected by phase noise in the second OPA, since coupling the small $e^{-2R_2}$ to the large  $e^{2R_2}$  will minimally affect the state. However, for small values of the squeezing parameter $R_2$, the phase noise $\theta_\text{rms,2}$ will couple in a similar way as $\theta_\text{rms,1}$ in equation \ref{eq:VeffTheta1}. In this scenario the second OPA will rotate the state, without providing sufficient anti-squeezing to neglect the effect of phase noise. 

Panel c and d of figure \ref{fig:ContourCascade} show the effective squeezing as a function of losses after the second OPA and phase noise in both the first and second OPA, respectively. The phase noise in the first OPA limits the achievable squeezing levels. This limitation becomes more pronounced when anti-squeezing is applied in the second OPA and the losses are mitigated. As the effect of losses becomes less critical, phase noise emerges as the dominant factor. The phase noise in the second OPA has little impact on the overall squeezing levels. This is due to the fact that while squeezing is highly sensitive to phase noise, anti-squeezing remains largely unaffected.

\section{Summary and conclusions}
We have proposed and analysed a single pass waveguide based squeezed-light source as a robust and scalable alternative to cavity based counterparts for gravitational wave detectors. The study used realistic parameters taken from \cite{kashiwazakiOver8dBSqueezedLight2023}, and looked at the effects of different types of losses, phase noise, and leakage. Furthermore, the cascaded OPA scheme was looked at in the context of reducing detection and out-coupling losses. Analytical equations were derived showing the insensitivity of phase noise in the second OPA when operating in the high gain regime. The models and results from the study can be used for all types of single-pass waveguides. Including larger weakly guiding waveguides and miniaturised on-chip waveguides. 

A waveguide based squeezer has the potential to directly address the phase noise limitations that limit squeezing in the current generation of gravitational wave detectors. In existing implementations, the dominant degradation arises from residual squeezing angle fluctuations caused by cavity length control noise, pump phase instability, and optical alignment drift. A waveguide squeezer can be expected to exhibit lower intrinsic phase noise susceptibility in comparison to its cavity counterparts owing to its reduced free space paths and compact, mechanically stable integration. While a direct, platform to platform, measurement of phase noise under identical conditions has not yet been published, this paper is the first step towards this analysis. Furthermore, waveguide squeezers are also expected to have greater resilience to gain induced diffraction  and backscatter. A full-scale implementation of this source could reduce operational complexity while improving the squeezer duty cycle, thereby laying the foundation for quantum noise mitigation in future, more sensitive detectors such as the Einstein Telescope.

\section*{Code availability}
All code from this study is available on GitHub at \cite{Svanberg_Code_availability_2025}.

\begin{acknowledgments}
This research was supported by the Swedish Research Council (VR starting grant 2023-0519), G{\"o}ran Gustafsson Prize for early career researchers and the Wallenberg Center for Quantum Technology (WACQT) in Sweden. The authors would also like to thank Terry McRae, Sheon Chua, Karmeng Kwan for useful discussions. We would also like to thank Katia Gallo and the support from the Optical Quantum Sensing environment grant 2016-06122. 
The authors declare no competing interests.
\end{acknowledgments}

\begin{appendix}
    \section{Matrix formalism} \label{app:Matrix}

    To facilitate easier numerical and analytical calculations, we employ a matrix formalism that models state transformations via transfer functions. Furthermore, the noise quadratures are represented in vector form as:

\begin{equation}
    \delta \vec{\hat{X}} =  \begin{pmatrix}
        \delta \hat{X}_{-} \\
        \delta \hat{X}_{+}
    \end{pmatrix} = \begin{pmatrix}
        -i & i \\ 1 & 1
    \end{pmatrix} \begin{pmatrix}
        \delta \hat{a} \\
        \delta \hat{a}^\dagger
    \end{pmatrix}
\end{equation}

Here $\delta \hat{a} $ represents the field fluctuations and $\delta \hat{X}_{\mp}$ represents the phase and amplitude quadratures. Continuing, we define matrices for the waveguide OPAs effect on the quadratures as: 

\begin{align}
    &M^{1} = \begin{pmatrix}
        \sqrt{\eta_{1\rightarrow 2}} e^{-R_1} & 0 \\ 0 & \sqrt{\eta_{1\rightarrow 2}} e^{R_1}
    \end{pmatrix} \\
    & M^{2} = \begin{pmatrix}
        \sqrt{\eta_{2\rightarrow \text{det}}} e^{R_2} & 0 \\ 0 & \sqrt{\eta_{2\rightarrow \text{det}}} e^{-R_2}
    \end{pmatrix} 
\end{align}

where $R_{1,2}$ represents the squeezing parameter of the first and second OPA respectively. Note here the switched sign in the exponent, indicating that the second OPA is anti-squeezing  with respect to the first in the phase quadrature. We also include the effects of phase noise with the rotation matrix:

\begin{equation}
    R_i = R(\theta_{\text{rms},i}) = \begin{pmatrix}
        \cos(\theta_{\text{rms},i}) & \sin(\theta_{\text{rms},i}) \\
        -\sin(\theta_{\text{rms},i}) & \cos(\theta_{\text{rms},i})
    \end{pmatrix}
\end{equation}

The state after the first OPA then becomes:

\begin{equation}
    \delta \vec{\hat{X}}^\text{out,1}=   \vec{TF}^\text{OPA,1}_\text{in} \delta \vec{\hat{X}}^\text{in} + \vec{TF}^\text{OPA,1}_{L,1} \delta \vec{\hat{X}}^{L,1}
\end{equation}

where $\delta \vec{\hat{X}}^\text{out,1}$, $\delta \vec{\hat{X}}^\text{in}$, and $\delta \vec{\hat{X}}^{L,1}$ are the noise quadratures of the output, input, and loss respectively and the transfer functions are:

\begin{align}
     & \vec{TF}^\text{OPA,1}_\text{in} = R_1 M^{1} R^{-1}_1 \\
    & \vec{TF}^\text{OPA,1}_{L,1} = \sqrt{1-\eta_{1\rightarrow 2}} \vec{I}
\end{align}

The second OPA will now amplify the entire previous state, this gives the state after the second OPA as:

\begin{equation}
    \delta \vec{\hat{X}}^\text{out}=   \vec{TF}^\text{amp}_\text{in} \delta \vec{\hat{X}}^\text{in} + \vec{TF}^\text{amp}_{L,1} \delta \vec{\hat{X}}^{L,1} + \vec{TF}^\text{amp}_{L,2} \delta \vec{\hat{X}}^{L,2} \label{eq:CascadedNoiseQuadratures}
\end{equation}

The transfer functions are now:

\begin{align}
    & \vec{TF}^\text{amp}_\text{in} = R_2 M^{2} R_1 M^{1} R^{-1}_1 R^{-1}_2 \\
    & \vec{TF}^\text{amp}_{L,1} = \sqrt{1-\eta_{1\rightarrow 2}} R_2 M^{2}R^{-1}_2 \\
    & \vec{TF}^\text{amp}_{L,2}  = \sqrt{1-\eta_{2\rightarrow \text{det}}} \vec{I}
\end{align}

By assuming all input and loss fluctuations are uncorrelated vacuum fluctuations, we get the variance from the noise quadrature in equation \ref{eq:CascadedNoiseQuadratures} by performing element wise squaring of the individual terms of the transfer functions and multiplying by the vacuum variance vector $(1~1)^{\mkern-1.5mu\mathsf{T}}$.
\end{appendix}

\bibliography{bib}
\end{document}